\documentstyle[11pt,moriond,epsfig]{article}
\def\QCD{\Lambda_{\overline{MS}}}
\def\OMT{ < \!\!  1-T \!\! > }
\def\MH{ < \!\! m_H^2/s \!\! > }
\begin{document}
\vspace*{4cm}
\title{Jet Physics: Theoretical Overview}

\author{ E.~W.~N.~Glover }

\address{Department of Physics, University of Durham,
Durham DH1 3LE, England}

\maketitle\abstracts{
I review the status of fixed order jet Monte Carlos and
briefly discuss the prospects for next-to-next-to-leading order
calculations.
I present a general purpose next-to-leading order
Monte Carlo program for
four jet event shape observables in
electron-positron annihilation and present some estimates
of the light jet mass and narrow jet broadening distributions.
Finally, I discuss an estimate of the strong coupling constant
using the measured energy evolution of the average value of
event shape variables such as thrust and heavy jet mass.}

\section{Introduction and status of fixed order parton level calculations}

In the last decade there has been enormous progress in using perturbative QCD
to predict and describe events containing jets.
At the simplest lowest-order (LO) level,
each jet is the footprint of a hard, well separated
parton produced in the event.
Although the predicted rate is sensitive to the choices of
renormalization and factorization scales, qualitative comparisons of
data and theory are generally very good.
Techniques for computing tree level multiparton scattering amplitudes
are very advanced and reliable numerical Monte Carlo programs exist
(see Table~\ref{program}) for processes such as
$p\bar p \to W+ 4$~jets~\cite{vecbos} which provides one of
the main backgrounds for
top quark detection at the TEVATRON.
However, the bulk of events populate topologies with fewer jets and
here a more quantitative description is required.  To date, this has been
largely achieved by improving the theoretical prediction to next-to-leading
order (NLO).

The addition of NLO effects produces four important
improvements over a LO estimate. First, the dependence
on the unphysical renormalization and factrorization scales
is reduced so that the
normalization is more certain.  Second, we begin to reconstruct the
parton shower so that two partons may combine to form a
single jet.  As a result, jet cross sections become sensitive to the
details of the jet finding algorithm, particularly the way in which
the hadrons are combined to form the jet axis and energy, and to the
size of a jet cone.  This sensitivity is also seen in experimental
results.  Third, the calculation becomes more sensitive to detector
limitations because radiation outside the detector is simulated. This
can change leading order results considerably for quantities such as
the missing transverse energy in events containing a $W$ boson.
Fourth,  the presence of infrared logarithms is clearly seen and regions where
resummations are needed to improve the perturbative prediction can be
identified.

Where possible, NLO estimates of QCD cross sections have become {\em de rigeur}
and have allowed many tests of the underlying dynamics and measurements
of, for example, the running strong coupling constant.
For all types of experiment, NLO Monte Carlos are playing a vital role
in making detailed comparisons with hadronic events.
Some examples of NLO programs are listed in Table~\ref{program}.
Recent examples where disagreement between theory and experiment have
yielded great excitement are the single jet inclusive data at CDF and
the $W+ 1$~jet/ $W+0$~jet ratio measured by D0.
For updates on these issues, I refer you to the talks by Elvira~\cite{elvira}
and Summers~\cite{summers}.

However, one loop matrix elements rapidly become more difficult to compute
as the number of external particles increases.
In the last few years, a number of problems connected with one-loop integrals
with five external particles have been solved and the one-loop
scattering amplitudes for five parton  \cite{5par}
and four partons and a vector boson \cite{us,them}
have been computed.
These matrix elements are relevant for
$p\bar p \to 3$~jets, $e^+e^- \to 4$~jets, $ep \to 3 + 1$~jets and
$p\bar p \to V + 2$~jets.
This latter process is likely to be very relevant at the TEVATRON in Run II,
where associated Higgs production $p\bar p \to V + H$ followed by
$H \to b\bar b$ is an eagerly anticipated discovery channel for the Higgs
boson.
NLO Monte Carlo programs for some of these processes are starting to appear
\cite{kilgore,trocsanyi} and \cite{menlo-parc,debrecen} (see
Table~\ref{program})
and comparisons with experimental data are underway.

However, while the NLO program has been extremely successful in
reducing the theoretical uncertainty, the improvements in the data
are even more impressive, to the extent that the theoretical error
tends to dominate.  For example, in extracting the strong coupling from
event shapes at LEP, the renormalization scale uncertainty engenders
approximately a 5\% uncertainty in $\alpha_s(M_Z)$ while the
experimental error is typically 2\%.
One way to improve the theoretical predictions is to incorporate
next-to-next-to-leading order (NNLO) effects.
This would reduce the
renormalization scale uncertainty and also allow a better interface
between the theoretical and experimental jet finding algorithms - now three
partons will be able to merge to form the jet.
Although a complete NNLO jet calculation is some way off, and for
$2 \to 2$ processes the two-loop double box integrals are not even known yet,
some encouraging steps have been taken in this direction.
For example, the singularity structures when two partons are unresolved has
been examined \cite{twounres}.
More impressive is the recent calculation of the
two-loop four gluon scattering amplitudes in a $N=4$ supersymmetric
version of QCD \cite{twoloop}.
While this result may not be directly relevant, the techniques developed to
obtain it undoubtedly will be and I expect rapid progress in the
evaluation of two loop QCD matrix elements.
Much work will still be needed to turn the amplitudes into cross sections,
but I predict that numerical NNLO programs will be available for $p\bar p \to
\leq 2$~jets and $e^+e^- \to 3$~jets will be available within the next five
years.

In the remainder of this talk, I will focus on two recent examples
of theoretical progress in jet physics.   First, I will discuss some new
results for four jet event shape observables using the recent
one loop matrix element calculations for $V \to 4~$partons \cite{us,them}.
Second, I will discuss how the theoretical error
on the strong coupling may be reduced by considering average values of
event shapes at a range of different centre-of-mass energies.

\begin{table}[t]
\begin{center}
\begin{tabular}{|c|c|c|c|}
\hline
Order & External legs & Physical process & Program \\
\hline \hline
LO   & $ \leq 7 $~partons     & $p\bar p \to \leq 5$~jets     &
{\tt NJETS} \cite{njets} \\
LO  & $ V + \leq 6 $~partons & $p\bar p \to W + \leq 4$~jets &
{\tt VECBOS} \cite{vecbos} \\
LO  & $ V + \leq 6 $~partons & $p\bar p \to Z + 4$~jets      &
\cite{barger} \\
LO  & $ V + \leq 6 $~partons & $e^+e^- \to 6$~jets           &
\cite{moretti} \\ \hline \hline
NLO &  4~partons & $p\bar p \to \leq 2$~jets & {\tt EKS} \cite{eks}, {\tt
JETRAD} \cite{dyrad} \\
NLO &  $V$ + 3~partons & $e^+e^- \to 3$~jets & {\tt EVENT} \cite{event},
{\tt EVENT2} \cite{event2} \\
NLO &  $V$ + 3~partons & $p\bar p \to V + \leq 1$~jet & {\tt DYRAD}
\cite{dyrad}
\\
NLO &  $V$ + 3~partons & $e p \to 2 + 1$~jet & {\tt DISENT} \cite{event2},
{\tt MEPJET} \cite{mepjet}, {\tt DISASTER} \cite{disaster}
\\
NLO &  5~partons & $p\bar p \to 3$~jets & \cite{kilgore}, \cite{trocsanyi}
\\
NLO &  $V$ + 4~partons & $e^+e^- \to 4$~jets & {\tt MENLO PARC}
\cite{menlo-parc}, {\tt DEBRECEN} \cite{debrecen}
\\
\hline
\end{tabular}
\caption{Available multi parton scattering matrix elements and
some commonly used fixed order Monte Carlo implementations.}
\label{program}
\end{center}
\end{table}

\section{4 jet observables}

As mentioned earlier, the one-loop matrix elements
appropriate for the decay of a vector boson into four partons have
recently been computed by two groups
\cite{us,them}.
These matrix elements have been combined with the
tree level amplitudes for the decay of a vector boson into five partons
to compute the NLO corrections to a variety
of four jet rates and event shapes by two groups;
the programs MENLO~PARC by Dixon and Signer \cite{menlo-parc} and
DEBRECEN by Trocsanyi and Nagy \cite{debrecen}.
Here we report on results obtained using
a third numerical implementation of these matrix elements
to compute infrared safe four jet observables, EERAD2 \cite{eerad2}.
This program uses the `squared' one-loop matrix elements of
\cite{us} together with squared tree level matrix elements for
$\gamma^* \to 5$~partons.
Both four and five parton contributions are singular in the
infrared limit and
there are several ways of performing the cancellation
\cite{slice,ert,event2,subtract}.
We use a modified version of the generic slicing approach
\cite{slice} that does not depend on the
slicing parameter $y_{\min}$.

As a check of the numerical results, Table~\ref{table} shows the
predictions for each of the three Monte Carlo programs for the four jet rate
for three jet clustering algorithms;
the Jade-E0,  Durham \cite{durham},  and Geneva \cite{geneva}
algorithms.
We show results with $\alpha_s(M_Z)=0.118$
for three values of the jet resolution parameter $y_{cut}$.
There is good agreement with the results from the other two calculations.

\begin{table}[t]
\begin{center}
\begin{tabular}{|c|c|c|c|c|}
\hline
Algorithm & $y_{{\rm cut}}$ & MENLO PARC & DEBRECEN & EERAD2 \\ \hline
\hline
         & 0.005& $ (1.04\pm0.02) \cdot 10^{-1} $ & $ (1.05\pm0.01) \cdot
10^{-1} $ & $ (1.05\pm0.01) \cdot 10^{-1} $\\
Durham & 0.01 & $ (4.70\pm0.06) \cdot 10^{-2} $ & $ (4.66\pm0.02) \cdot
10^{-2} $ & $ (4.65\pm0.02) \cdot 10^{-2} $\\
       & 0.03 & $ (6.82\pm0.08) \cdot 10^{-3} $ & $ (6.87\pm0.04) \cdot
10^{-3} $ & $ (6.86\pm0.03) \cdot 10^{-3} $\\
\hline
\hline
       & 0.02 & $ (2.56\pm0.06) \cdot 10^{-1} $ & $ (2.63\pm0.06) \cdot
10^{-1} $ & $ (2.61\pm0.05) \cdot 10^{-1} $\\
Geneva & 0.03 & $ (1.71\pm0.03) \cdot 10^{-1} $ & $ (1.75\pm0.03) \cdot
10^{-1} $ & $ (1.72\pm0.03) \cdot 10^{-1} $\\
       & 0.05 & $ (8.58\pm0.15) \cdot 10^{-2} $ & $ (8.37\pm0.12) \cdot
10^{-2} $ & $ (8.50\pm0.06) \cdot 10^{-2} $\\
\hline
\hline
       & 0.005& $ (3.79\pm0.08) \cdot 10^{-1} $ & $ (3.88\pm0.07) \cdot
10^{-1} $ & $ (3.87\pm0.03) \cdot 10^{-1} $\\
JADE-E0 & 0.01& $ (1.88\pm0.03) \cdot 10^{-1} $ & $ (1.92\pm0.01) \cdot
10^{-1} $ & $ (1.93\pm0.01) \cdot 10^{-1} $\\
       & 0.03 & $ (3.46\pm0.05) \cdot 10^{-2} $ & $ (3.37\pm0.01) \cdot
10^{-2} $ & $ (3.35\pm0.01) \cdot 10^{-2} $\\
\hline
\end{tabular}
\caption{The four-jet fraction as calculated by MENLO~PARC,
DEBRECEN and EERAD2,
for the different jet recombination schemes and varying $y_{\rm cut}$.
The rate is normalized by the ${\cal O}(\alpha_s)$
total hadronic cross-section, $\sigma_{\rm tot}=\sigma_0 \, (1+
\alpha_s/\pi)$.}
\label{table}
\end{center}
\end{table}

In addition to the jet rates, there are also NLO
predictions in the literature
for four-jet event shapes such as the $D$ parameter, Acoplanarity and
the Fox-Wolfram moments \cite{debrecen}.
These variables vanish as the three jet limit approaches;  i.e. as the
event becomes more planar.
Of course there are many other event shape variables that vanish
as the event becomes more three jet like.
Here, I focus on observables derived by dividing the event into two
hemispheres $H_1$ and $H_2$
according to the orientation of the thrust axis $\vec{n}$
defined by,
\begin{equation}
T = \max \frac{\sum_k | \vec{p}_k.\vec{n} |}{\sum_k |\vec{p}_k|}.
\end{equation}
Particles that satisfy $\vec{p}_i.\vec{n} > 0$ are assigned to hemisphere
$H_1$, while all other particles are in $H_2$.
The light hemisphere mass, $m_L^2/s$, and the narrow jet broadening,
$B_{\min}$, are defined by,
\begin{eqnarray}
\frac{m_L^2}{s} &=& \frac{1}{s}  \min_{i=1,2} \left(\sum_{k\in H_i}
p_k\right)^2\nonumber \\
B_{\min} &=& \min_{i=1,2} \frac{\sum_{k\in H_i} |\vec{p}_k\times \vec{n}|}
{2 \sum_k |\vec{p}_k | },
\label{eq:defs}
\end{eqnarray}
are both non-zero when there are at least four particles in the final state.
These variables are related to
the heavy jet mass $m_H^2/s$ and the wide jet broadening $B_{\max}$
which are obtained by maximising the quantities in Eq.~\ref{eq:defs},
and which require at least three particles in the event to be non-zero.

\begin{figure}[t]
\begin{center}
\psfig{figure=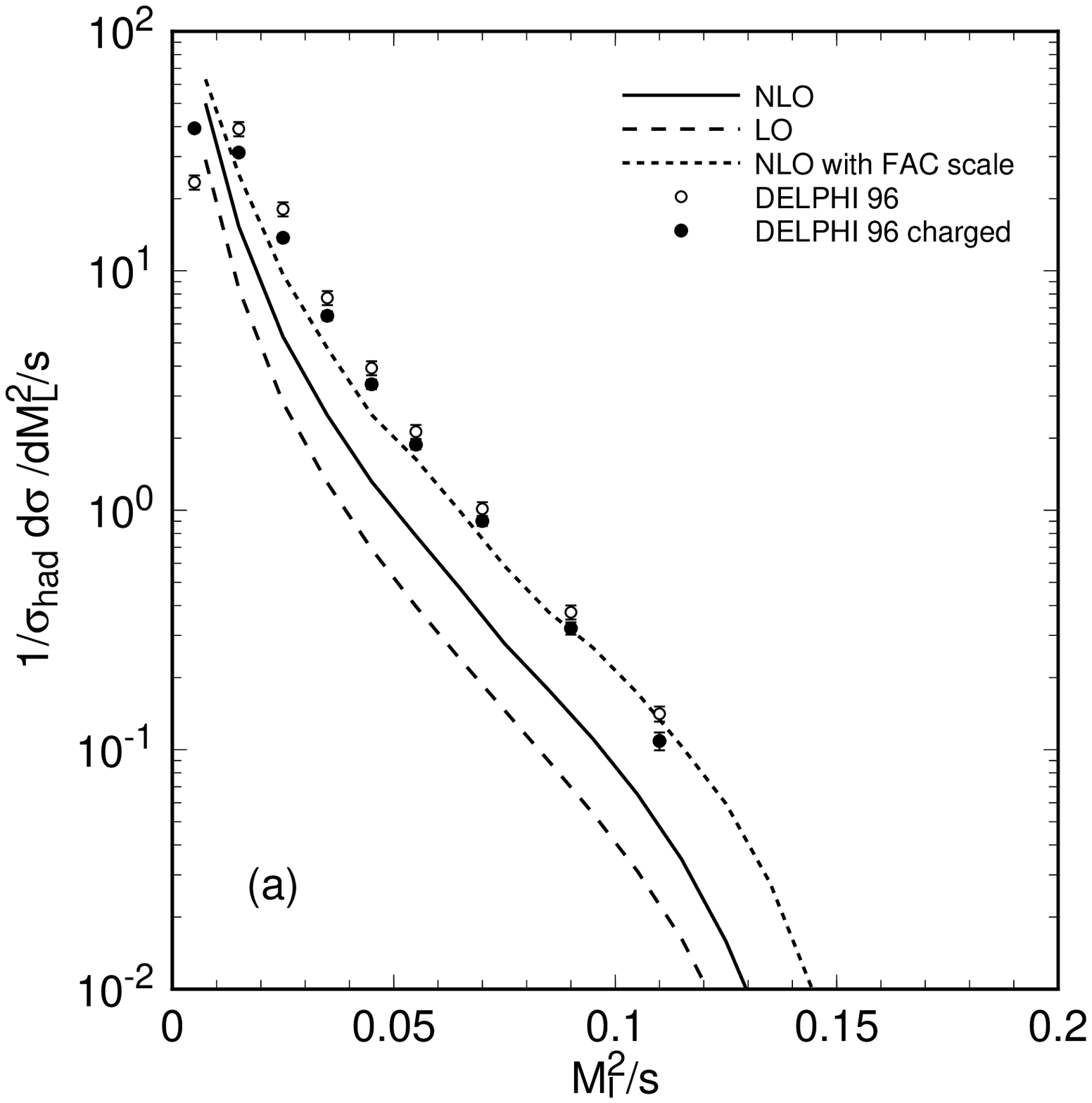,width=6cm}
\psfig{figure=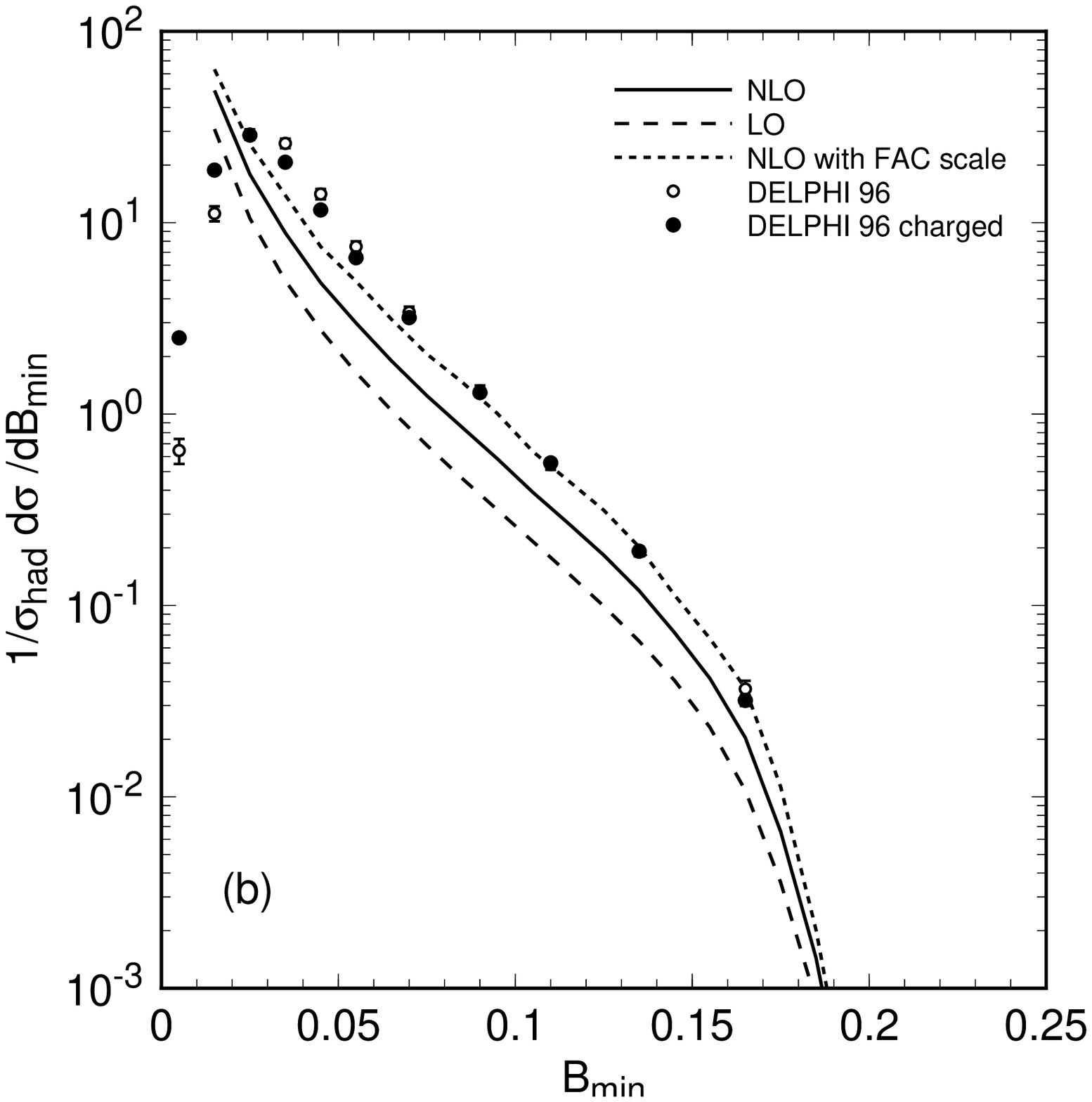,width=6cm}
\end{center}
\caption[]{The (a) light hemisphere mass distribution
$1/\sigma_{had} d\sigma / d (M_L^2/s)$ and (b)
narrow jet broadening distribution
$1/\sigma_{had} d\sigma / d B_{\min}$ at
LEP energies.
The LO (NLO) prediction is shown dashed (solid) for
$\mu = M_Z$.   The LO (and NLO) prediction using the FAC scale
is shown as a dotted line.
The data is taken from \cite{4jetdata}.}
\label{fig:4shape}
\end{figure}

The LO and NLO predictions for
$m_L^2/s$ and $B_{\min}$ with a renormalization scale
$\mu = M_Z$ and $\alpha_s(M_Z) = 0.1194$
are shown in fig.~\ref{fig:4shape},
together with data from \cite{4jetdata}.
At small values of the observable, we see the presence of
large logarithms which must be resummed to obtain a meaningful result.
At larger values,
we see that the next-to-leading order corrections are large and
approximately 100\%.
Similar large effects have been noted for other four jet observables
\cite{menlo-parc,debrecen}.
While such large effects may make the perturbation theory appear unreliable,
we note that effects of similar size have been noted
for three jet event shapes
like thrust \cite{ert,event}.
This may be a sign that large higher order
or power corrections are present.
Or it may simply indicate that the physical scale $\mu = M_Z$ is a very
poor choice of the renormalization scale, which itself engenders large
ultraviolet
logarithms in the higher order terms.
In fact, such ultraviolet logarithms may be resummed using the
FAC or fastest apparent convergence scale \cite{grunberg}.
This scale is usually much less than the physical scale (and particularly
so when the next-to-leading order corrections appear large) and may be
disfavoured for that reason.   However, the prediction using this scale
(shown as dotted lines in fig.~\ref{fig:4shape}.)
does lie much closer to the data.

\section{Strong coupling from the energy evolution of event shapes}

A vital ingredient in perturbative predictions is an accurate knowledge
of the strong coupling constant.  This can be determined via analysis of event
shape variables at LEP.
Consider the observable $R(Q)$ with
a perturbation series and leading power correction that describes the
hadronization phase of the hard scattering of the form,
\begin{equation}
R(Q)=a+r_1 a^2+r_2 a^3 + \ldots +
\frac{\lambda}{Q} (1+\lambda_1a + \ldots),
\label{eq:thrust}
\end{equation}
where $a \equiv \alpha_S(\mu)/\pi$ denotes the renormalization
group improved coupling.  Note that the normalization is simply such
that the perturbative expansion begins with unit coefficient.
An example of such a variable is $R(Q) = \OMT/1.05$,
where in the
${\overline{MS}}$ scheme with $\mu=Q$ and $N_f=5$ active quark
flavours the NLO coefficient is $r_1 =
9.70$~\cite{ert,event}. The NNLO
coefficient $r_2$ is as yet unknown.   The precise form of the power
corrections is as yet not fully understood, but, for the purposes of comparison
with data, may be parameterized in a variety of different ways \cite{wicke}.
To extract $\alpha_s(M_Z)$ from the data,
we just truncate the perturbative series for a given renormalization
scale $\mu = xQ$ (which is typically $x=1$).  In other words, we assume the
higher order term $r_2\equiv 0$, $r_3 \equiv 0$ etc.
Then, by comparing with experimental data,
we solve for $a$.  A recent analysis \cite{siggi} for $\OMT$ using
the power correction model of \cite{webber} finds,
\begin{equation}
\alpha_s(M_Z) = 0.1204 \pm 0.0013 \phantom{~}^{+0.0061}_{-0.0050} \phantom{~}
^{+0.0023}_{-0.0018},
\label{eq:sigresult}
\end{equation}
(with a $\chi^2/{\rm d.o.f}$ of $42.6/24$) where the first error is
purely experimental.  The second and third errors come from varying
the theoretical input parameters, first varying the renormalization scale
$x$ between 0.5 and 2
and second the parameters of the power correction model.  Clearly the estimate
of the theoretical error is dominated by the renormalization scale
uncertainty.
Similar results are presented in \cite{wicke}.

Alternatively, we may avoid the renormalization scale entirely and
directly write an expression for the running of $R(Q)$ with $Q$ in
terms of $R(Q)$ itself~\cite{grunberg,ECother},
\begin{eqnarray}
\frac{dR}{d\log Q} &=& -b R^2(1+cR+\rho_2 R^2 + \ldots )
                       +K_0 R^{-c/b} e^{-1/bR}
                        (1+K_1 R + \ldots) +\ldots \nonumber \\
                   &\equiv& -b \rho(R).
\label{eq:running}
\end{eqnarray}
Here $b$ and $c$ are the first two universal terms of the
QCD beta-function,
\begin{equation}
b=\frac{33-2N_f}{6},\qquad\qquad
c=\frac{153-19N_f}{12b}.
\end{equation}
The quantity,
\begin{equation}
\rho_2 \equiv r_2+c_2-r_1c-r_1^2,
\label{eq:r2}
\end{equation}
is a renormalization scheme and renormalization scale (RS) invariant
combination of the NLO and NNLO perturbation series and beta-function
coefficients with, in the $\overline{MS}$ scheme,
\begin{equation}c_2 = \frac{77139-15099N_f+325N_f^2}{1728b}.\end{equation}
Since the NNLO $r_2$ is unknown, so is $\rho_2$.
The coefficient
$K_0$ is directly related to the coefficient $\lambda$ of the $1/Q$ power
corrections in eq.~(\ref{eq:thrust}).

Since $R(Q)$ and $dR/d\log Q$ are both observables one could
in principle directly fit eq.~(\ref{eq:running}) to the data and thus
constrain the unknown coefficients $\rho_2$ and $K_0$. At asymptotic
energies all observables run universally according to,
\begin{equation}
\frac{dR}{d\log Q} = -bR^2(1+cR),
\label{eq:univrun}
\end{equation}
and one could see how close the data are to this evolution
equation. Given the error bars of the data and the separation in $Q$
of the different experiments it is preferable, however, to integrate
up eq.~(\ref{eq:running}) using asymptotic freedom ($R(Q) \to 0$ as $Q
\to \infty$) as a boundary condition.  In this way one
obtains,
\begin{equation}
\frac{1}{R }+c \log \left( \frac{cR}{1+cR} \right)
 = b \log\left(\frac{Q}{\Lambda_R}\right)
  - \int_0^{R} dx \, \left(-\frac{1}{\rho(x)}+\frac{1}{x^2(1+cx)} \right),
\label{eq:intup}
\end{equation}
where $\Lambda_R$ is a constant of integration. By comparing with the
$Q \to
\infty$ behaviour of eq.~(\ref{eq:thrust}) one can deduce
that $\Lambda_R$ is related to $\QCD$,
\begin{equation}
\Lambda_R = e^{r/b} \left(\frac{2c}{b}\right)^{-c/b} \QCD,
\label{eq:RtoMS}
\end{equation}
where $r \equiv r_1^{\overline{MS}}(\mu = Q)$.

{\bf If} we assume that the right-hand side of eq.~(\ref{eq:running}) is
adequately parameterized by,
\begin{equation}
-b\rho(R)=-bR^2(1+cR+\rho_2R^2)+K_0 R^{-c/b}e^{-1/bR},
\label{eq:param}
\end{equation}
we can then insert this form into eq.~(\ref{eq:intup}) and by
(numerically) solving the transcendental equation, perform fits of
$\rho_2$, $K_0$ (or equivalently $\lambda$) and $\QCD$ to the data $R(Q)$.

\begin{figure}[t]
\begin{center}
\psfig{figure=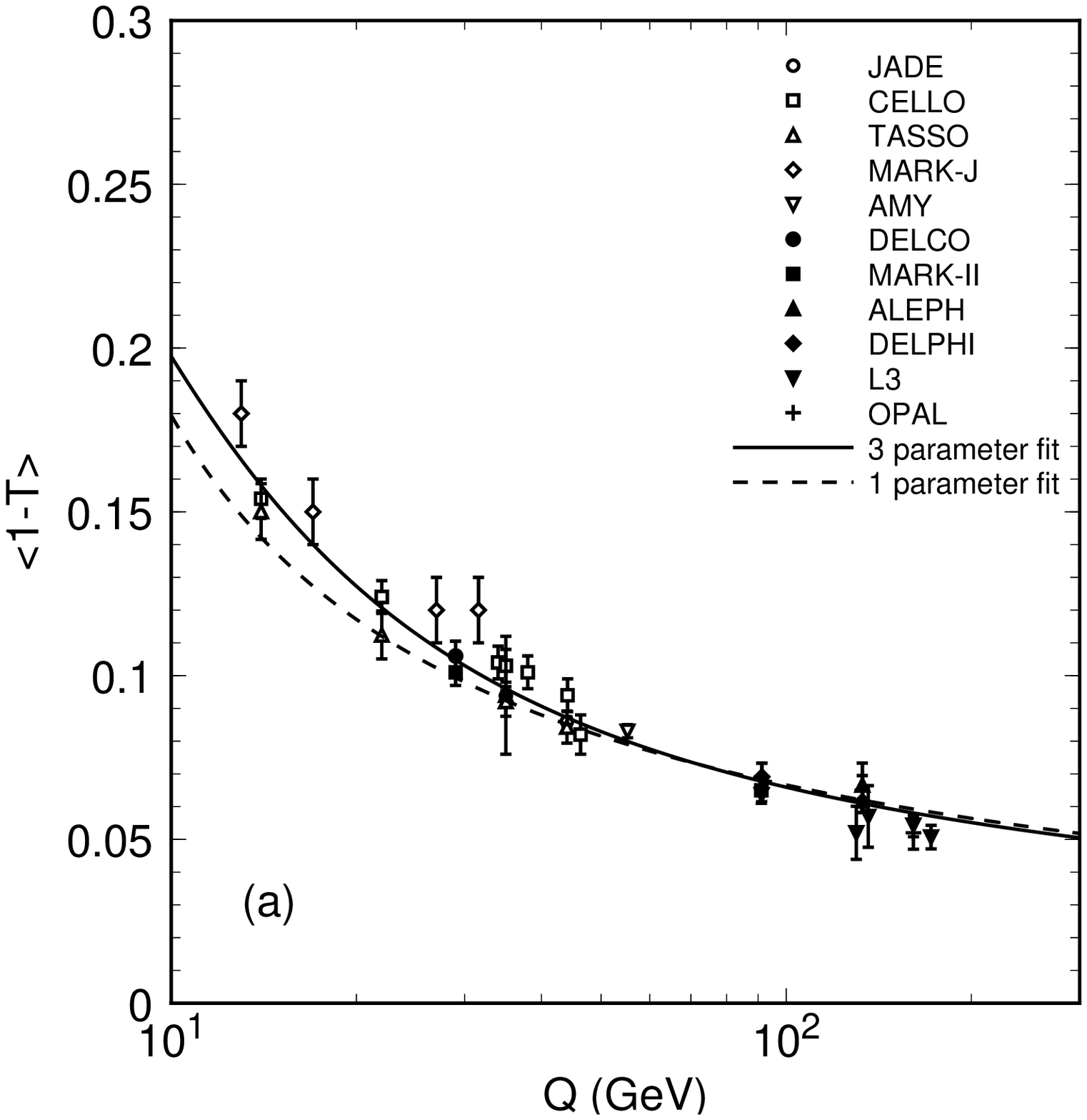,width=6cm}
\psfig{figure=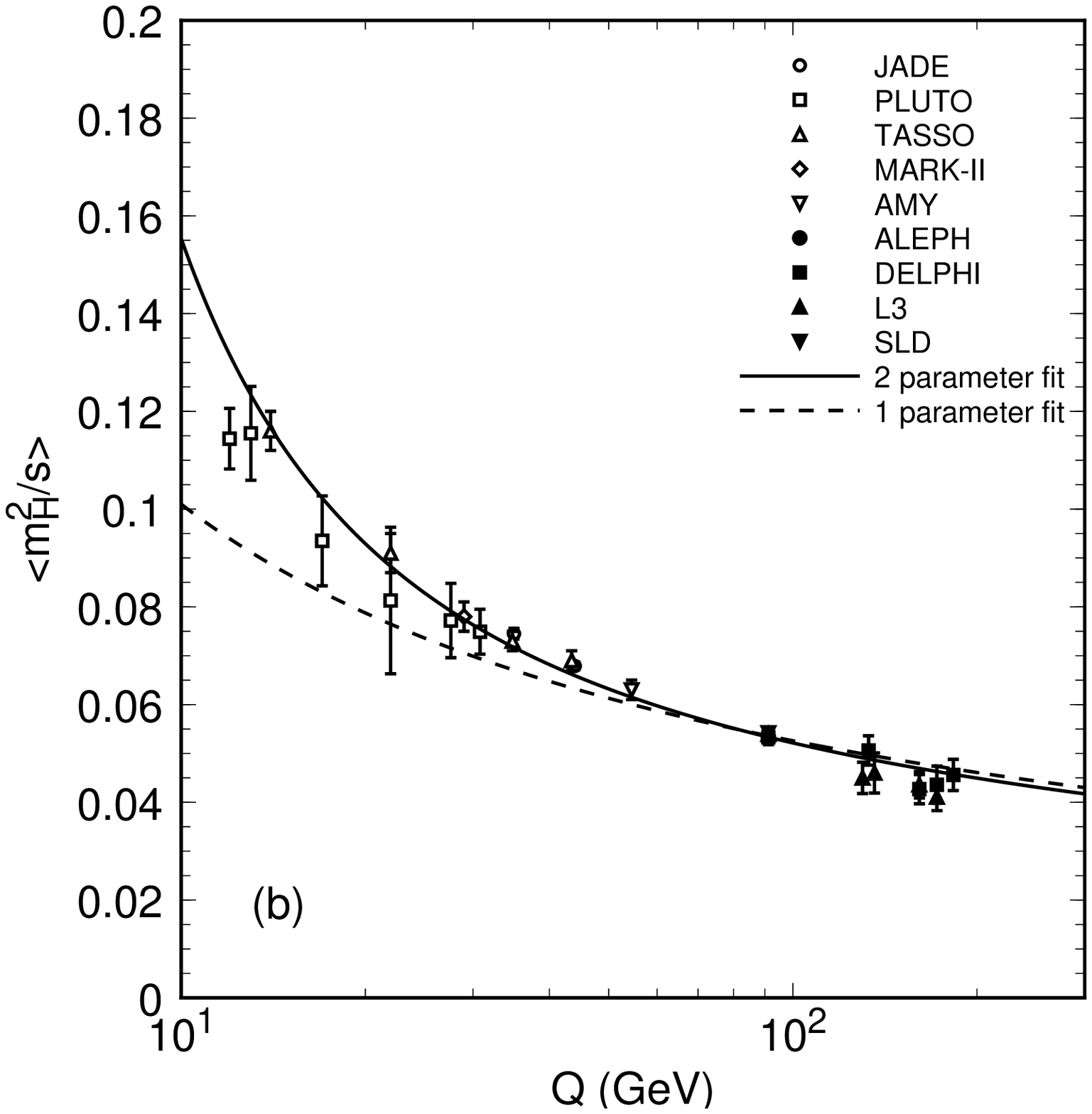,width=6cm}
\end{center}
\caption[]{The average value of (a) 1-Thrust and (b) heavy jet mass
obtained experimentally
compared with the expectation of eq.~(\ref{eq:intup}).
The dashed lines show the fit to the data with $\rho_2 = \lambda = 0$
while in (a) the result of the three parameter fit (to $\QCD, ~\rho_2$ and
$\lambda$)
is shown as a solid line. In (b), the solid line shows the
result of a two parameter fit (to $\rho_2$ and $\lambda$) using the
value of $\QCD$ obtained from the three parameter fit to $\OMT$. }
\label{fig:3parfit}
\end{figure}

Fig.~\ref{fig:3parfit}(a) shows the fit to the data (dashed line)
obtained by
setting $\rho_2=\lambda=0$. This corresponds to the universal running
of the observable given in eq.~(\ref{eq:univrun}). The fitted value is
$\QCD^{(5)} = 266~{\rm MeV}$ with a $\chi^2/{\rm d.o.f} = 81.7 / 32$.
We clearly see that the data is
falling much too quickly with increasing $Q$ for the asymptotic
behaviour to have set in at these scales.  The data favours a more
steeply falling evolution which could be caused by either higher order
corrections with a positive $\rho_2$, or power corrections with
non-zero $K_0$.  We therefore perform a 3-parameter fit allowing
$\rho_2$, $K_0$ and $\QCD$ to vary independently which is shown as a
solid line in Fig.~\ref{fig:3parfit}(a).  The minimum
$\chi^2$ fit is acceptable ($\chi^2/{\rm d.o.f} = 40.4 / 30 $) and
estimating an error by allowing $\chi^2$ within $1$ of the minimum
gives,
\begin{displaymath}
\QCD^{(5)}=245 ^{+20}_{-17}~{\rm MeV}
\qquad {\rm with} \qquad
\rho_2=-16\mp 11
\qquad {\rm and} \qquad
\lambda=0.27^{+0.12}_{-0.10}~{\rm GeV}.
\end{displaymath}
These values of $\rho_2$ and $\lambda$ are reasonably small, thereby
lending support to our critical assumption that the evolution equation
could be parameterized in this way.  Converting the extracted
value of $\QCD$ into $\alpha_S(M_Z)$ we find,
\begin{equation}
\alpha_S(M_Z)=0.1194 \pm 0.0014.
\end{equation}

We see that our central value is remarkably close to that obtained by
\cite{siggi,wicke}.  The main difference is in how the errors are
determined.  In our approach, the errors are estimated by allowing the
uncalculated higher orders to be fitted by the data and the data
prefers these to be small.
In particular, the
renormalization group scale-dependent logarithms are automatically
resummed to all orders on integrating eq.~(\ref{eq:running}) and do
not add a spurious extra large uncertainty in the extraction of $\QCD$
(or equivalently $\alpha_s(M_Z)$).
As higher order corrections become known,
the new RS-invariant terms, $\rho_2$, $\rho_3$ etc., can be
incorporated and the fit refined.

With such an accurate value of $\alpha_S(M_Z)$, we should expect that
applying this approach to other variables should yield consistent results.
Unfortunately, the method described here relies on having
reliable data over a wide range of $Q$ values.
This limits its use to a very few variables,
like thrust or the heavy jet mass.
If we repeat the same analysis for the average of the
heavy jet mass, $\MH$, we find that
a one parameter fit with $\rho_2 = \lambda = 0$
gives a very poor fit, $\chi^2/{\rm d.o.f} = 213/29$ and $\QCD = 368$~MeV.
As seen in Fig.~\ref{fig:3parfit}(b)
the data evolves much faster than the QCD prediction.
However, allowing both $\rho_2$ and $\lambda$ to vary while using the
value of $\QCD = 245$~MeV obtained from the $\OMT$ analysis gives a much
more satisfactory description.\footnote{Unfortunately,
in a three parameter fit,
$\rho_2$ and $\lambda$ trade off against each other and drive $\rho_2$ and
$\lambda$ to unacceptably large values
where we would have no reason to believe that
the parameterization is adequate.}
Here, $\chi^2/{\rm d.o.f} = 40.3/27$ while $\rho_2 = 13$ and
$\lambda = 0.11$~GeV  are sufficiently
small to support our choice of parameterization.

\section{Outlook}
The last decade has seen enormous progress both experimentally and
theoretically in understanding high energy hadronic events.
In many cases, we now have a quantitative understanding of the rates
for strong interaction processes and detailed predictions for the
structure of the events.
One of the main theoretical uncertainties remains the
renormalization scale uncertainty and the value of the
strong coupling.  New techniques for
next-to-next-to-leading predictions are in
sight and will help to reduce this
error.  However, a more immediate improvement may be obtained
by resumming the ultraviolet
logarithms explicitly.

\section*{Acknowledgments}
I gratefully acknowledge the financial support of the Royal Society.
It is a pleasure to thank Chris Maxwell, John Campbell
 and Matt Cullen
for stimulating collaborations.


\section*{References}
\begin{thebibliography}{99}
\bibitem{vecbos}
{F.A. Berends, H. Kuijf, J.B. Tausk and W.T. Giele, Nucl. Phys. {\bf B357}
(1991) 32.}
\bibitem{elvira}
{V.D. Elvira, these proceedings.}
\bibitem{summers}
{D.J. Summers, these proceedings.}
\bibitem{5par}
{Z. Bern, L. Dixon and D.A. Kosower, Phys. Rev. Lett. {\bf 70} (1993) 2677;
Z. Kunszt, A. Signer and Z. Tr{\'o}cs{\'a}nyi, Phys. Lett. {\bf B336} (1994)
529;
Z. Bern, L. Dixon and D.A. Kosower, Nucl. Phys. {\bf B437} (1995) 259}
\bibitem{us}{
E.W.N. Glover and D.J. Miller, Phys. Lett. {\bf B396} (1997) 257;
J.M. Campbell, E.W.N. Glover and D.J. Miller, Phys. Lett. {\bf B409} (1997)
503.
}
\bibitem{them}{
Z. Bern, L. Dixon and D.A. Kosower, Nucl. Phys. Proc. Suppl. {\bf 51C} (1996)
243;
Z. Bern, L. Dixon, D.A. Kosower and S. Weinzierl, Nucl. Phys. {\bf B489} (1997)
3;
Z. Bern, L. Dixon and D.A. Kosower, Nucl. Phys. {\bf B513} (1998) 3
}
\bibitem{kilgore}
{W.B. Kilgore and W.T. Giele, Phys. Rev. {\bf D55} (1997) 7183;
W.B. Kilgore, in {\em Proceedings of 32nd Rencontres de Moriond; QCD and High
Energy Hadronic Interactions}, Les Arcs, France, March 1997, hep-ph/9705384.}
\bibitem{trocsanyi}
{Z. Tr{\'o}cs{\'a}nyi, Phys. Rev. Lett. {\bf 77} (1996) 2182.}
\bibitem{menlo-parc}
{A. Signer and L. Dixon, Phys. Rev. Lett. {\bf 78} (1997) 811;
A. Signer and L. Dixon, Phys. Rev.  {\bf D56} (1997) 4031;
A. Signer, Comput. Phys. Comm. {\bf 106} (1997) 125;
A. Signer, hep-ph/9705218.
}
\bibitem{debrecen}
{Z.~Nagy and Z.~Tr{\'o}cs{\'a}nyi, Phys. Rev. Lett. {\bf 79} (1997) 3604;
hep-ph/9708343;
hep-ph/9708344;
hep-ph/9712385.
}
\bibitem{twounres}
{F.A. Berends and W.T. Giele, Nucl. Phys. {\bf B313} (1989) 595;
S. Catani, Proceedings of Workshop on `New Techniques for
Calculating Higher Order QCD Corrections',
preprint ETH-TH/93-01, Zurich (1992);
J.M. Campbell and E.W.N. Glover, hep-ph/9710255;
S. Catani, hep-ph/9802439.
}
\bibitem{twoloop}
{Z. Bern, J.S. Rozowsky and B. Yan, Phys. Lett. {\bf B410} (1997) 273,}
\bibitem{njets}
{F.A. Berends, W.T. Giele and H. Kuijf, Phys. Lett. {\bf B232} (1989) 266:
F.A. Berends and H. Kuijf, Nucl. Phys. {\bf B353} (1991) 59.}
\bibitem{barger}{V. Barger, E. Mirkes, R.J.N. Phillips and T. Stelzer, Phys.
Lett. {\bf B338} (1994) 336.}
\bibitem{moretti}{S. Moretti, Phys. Lett. {\bf B420} (1998) 367.}
\bibitem{eks}
{S.D. Ellis, Z. Kunszt and D.E. Soper,
              Phys. Rev. {\bf D40}, 2188 (1989);
	      Phys. Rev. Lett. {\bf 64}, 2121 (1990);
              Phys. Rev. Lett. {\bf 69}, 1496 (1992).}
\bibitem{dyrad} {W.T. Giele, E.W.N. Glover and D.A. Kosower,
Nucl. Phys. {\bf B403} (1993) 633.}
\bibitem{event}{
Z. Kunszt, P. Nason, G. Marchesini and B.R. Webber,
in {\em Z Physics ar LEP 1}, vol. 1, ed. G. Altarelli, R. Kleiss and C.
Verzegnassi, CERN Yellow Report 89-08.
}
\bibitem{event2} {S. Catani and M.H. Seymour, Phys. Lett. {\bf B378} (1996)
287; Nucl. Phys. {\bf B485} (1997) 291. }
\bibitem{mepjet}{E. Mirkes and D. Zeppenfeld, Phys. Lett. {\bf B380} (1996)
205.}
\bibitem{disaster} {D. Graudenz, hep-ph/9708362.}
\bibitem{eerad2}
{J.M. Campbell, M. Cullen and E.W.N. Glover, in preparation}
\bibitem{slice}{
K. Fabricius, I. Schmitt, G. Kramer and G. Schierholz,
Z. Phys. {\bf C11} (1981) 315;
W.T. Giele and E.W.N. Glover,
Phys. Rev. {\bf D46} (1992) 1980.
}
\bibitem{ert}{
R.K. Ellis, D.A. Ross and A.E. Terrano,
Nucl. Phys. {\bf B178} (1981) 421.}
\bibitem{subtract}{
S. Frixione, Z. Kunszt and A. Signer, Nucl Phys. {\bf B467} (1996) 399;
Z. Nagy and Z. Tr{\'o}cs{\'a}nyi, Nucl. Phys. {\bf B486} (1997) 189.
}
\bibitem{durham}{
Yu.L. Dokshitzer, Contribution to the Workshop on Jets
at LEP and HERA, J. Phys. {\bf G17} (1991) 1441.}

\bibitem{geneva}
{S. Bethke, Z. Kunszt, D.E. Soper and W.J. Stirling, Nucl. Phys. {\bf B370}
(1992) 310. }

\bibitem{4jetdata}{P. Abreu et al, DELPHI Collaboration, Z. Phys. {\bf C73}
(1996) 11.}

\bibitem{wicke} {D. Wicke, these proceedings.}
\bibitem{ECother}{
A. Dhar and V. Gupta, Phys. Rev. {\bf D29} (1984) 2822;
D.T. Barclay, C.J. Maxwell and M.T. Reader,
Phys. Rev. {\bf D49} (1994) 3480;
C.J. Maxwell, Phys. Lett. {\bf B409} (1997) 450.
}
\bibitem{siggi}{
P.A. Movilla Fern{\'a}ndez et al and the JADE Collaboration, Eur. Phys. J. {\bf
C1} (1998) 461.
}
\bibitem{thrust}{J.M. Campbell, E.W.N. Glover and C.J. Maxwell,
hep-ph/9803254.}

\bibitem{webber}{
Yu.L. Dokshitzer and B.R. Webber,
Phys. Lett. {\bf B352} (1995) 451.}

\bibitem{grunberg}{
G. Grunberg, Phys. Lett. {\bf B95} (1980) 70; Phys. Rev. {\bf D29} (1984) 2315.
}

\end{thebibliography}
\end{document}